\documentclass[prl,twocolumn,bibnotes,showpacs,amsbsy,floatfix,amsbsy,floatfix,superscriptaddress]{revtex4-1}
\usepackage{epsfig,color}
\usepackage[cp1250]{inputenc}
\usepackage{amsmath}
\usepackage{amssymb}
\usepackage{amsfonts}
\usepackage{color}
\usepackage{wasysym}
\usepackage{latexsym} 
\usepackage[dvipsnames]{xcolor}

\usepackage{graphicx,color}

\begin{document}

\title{Hinge Modes and Surface States in Second-Order Topological Three-Dimensional \\ Quantum Hall Systems induced by Charge Density Modulation  }
\author{Pawe\l{} Szumniak}
\affiliation{AGH University of Science and Technology, Faculty of
Physics and Applied Computer Science,\\
al. Mickiewicza 30, 30-059 Krak\'ow, Poland}
\author{Daniel Loss}
\affiliation{Department of Physics, University of Basel, Klingelbergstrasse 82, 4056 Basel, Switzerland}
\author{Jelena Klinovaja}
\affiliation{Department of Physics, University of Basel, Klingelbergstrasse 82, 4056 Basel, Switzerland}
\date{\today}

%\pacs{71.10.Fd; % 	Lattice fermion models (Hubbard model, etc.)
%73.43.-f; %	Quantum Hall effects
%71.10.Pm %	Fermions in reduced dimensions (anyons, composite fermions, Luttinger liquid, etc.) (for anyon mechanism in superconductors, see 74.20.Mn)
%}
%73.43.Nq; %	Phase transitions: Quantum Hall effects, 
%73.63.Nm	Quantum wires 
%74.45.+c	Proximity effects; Andreev reflection; SN and SNS junctions
%71.10.Fd 	Lattice fermion models (Hubbard model, etc.) 
% 71.10.Pm 	Fermions in reduced dimensions (anyons, composite fermions, Luttinger liquid, etc.) (for anyon mechanism in superconductors, see 74.20.Mn)
%73.21.Cd 	Superlattices 
%73.21.Hb 	Quantum wires
% 73.21.La 	Quantum dots 
%05.30.Pr	Fractional statistics systems (anyons, etc.)
%73.43.Nq,	Phase transitions: Quantum Hall effects, 
%73.43.-f	Quantum Hall effects, 
% 85.35.Be	Low-dimensional structures: devices
% 73.63.Nm	electronic transport in Quantum wires
% 03.67.Lx	Quantum computation
%Charge-density waves
%71.45.Lr	collective excitations, 
%72.15.Nj	one-dimensional conductors, 
%73.20.Mf	surface and interface excitations	
%72.15.Nj %	Charge-density waves, one-dimensional conductors

\begin{abstract}
We consider a system of weakly coupled one-dimensional wires forming a three-dimensional stack in the presence of a spatially periodic modulation of the chemical potential along the wires, equivalent to a charge density wave (CDW). An external static magnetic field is applied parallel to the wire axes.  We show that, for a certain parameter regime, due to interplay between the CDW and magnetic field, the system can support a second-order topological phase characterized by the presence of  chiral quasi-1D Quantum Hall Effect (QHE) hinge modes. Interestingly, we demonstrate that direction of propagation of the hinge modes depends on the phase of the CDW and can be reversed only by electrical means without the need of changing the orientation of the magnetic field. Furthermore, we show that the  system can also support 2D chiral surface QHE states, which can coexist with one-dimensional  hinge modes, realizing a scenario of a hybrid high-order topology. We show that the hinge modes are robust against static disorder.

\end{abstract}

\maketitle

{\it Introduction.} 
Over the last decade  topological phases of matter have attracted considerable attention in condensed matter physics. This was triggered by the discovery of the quantum Hall effect (QHE)~\cite{Klitzing, Tsui,Girvin,Mcdonald} with striking stability of the edge states~\cite{TKNN}.
These findings motivated a large amount of experimental and theoretical work on other topological systems such as fractional QHE~\cite{Laughlin, Jain2}, topological insulators (TIs) and superconductors (TSCs). In $d$-spatial dimension, the bulk of those systems is gapped while there exist topologically protected gapless states  on their ($d-1$)-dimensional boundaries~\cite{Liang_Zhang_RMP}. 
Very recently concepts of topological materials have been generalized to a new class of $d$-dimensional systems which host topologically protected edge states on ($d-n$) dimensional boundaries, which are referred to as $n$-th order TIs and TSCs~\cite{Benalcazar2017,Benalcazar20172,Geier2018,Benalcazar,song2017,Peng2017,Imhof2017,Langbehn,Schindler2018,Hsu2018,Ezawa2018,Ezawa20182,Ezawa20183,Zhu2018,Wang2018,Zhang2018,Wang20182,Liu2018,Yan2018, Yanick_HOTSC,TB_class,Bultinck_hybrid_topology, HOTI_FCC, MCS_Hughes_2019,ortix}.

In the present work, we propose a novel 3D system related to the QHE and uncover  striking properties such as  1D chiral hinge modes - realizing second-order topological QHE and 2D chiral surface QHE states - present in 3D QHE~\cite{3DQH_1, 3DQH_2, 3DQH_3, 3DQH_4, 3DQH_5, 3DQH_6, 3DQH_7, 3DQH_8, 3DQH_9}, recently discovered in ZrTe$_5$ \cite{3DQH_exp}. Furthermore, the studied system can host both types of states coexisting in the gap supporting a hybrid scenario with mixed higher-order topology, investigated in the context of higher-order TSCs~\cite{Bultinck_hybrid_topology,ortix}.
Remarkably, the direction of propagation of the hinge modes can be tuned by changing the phase of the CDW potential. This extraordinary feature paves the way for an all-electrical control of the propagation direction  of  topological hinge modes without the need of changing the orientation of the external magnetic field, neither does it require  any spin-orbit interaction.

\begin{figure}[bt!]
\centering
\includegraphics[width=8.6cm]{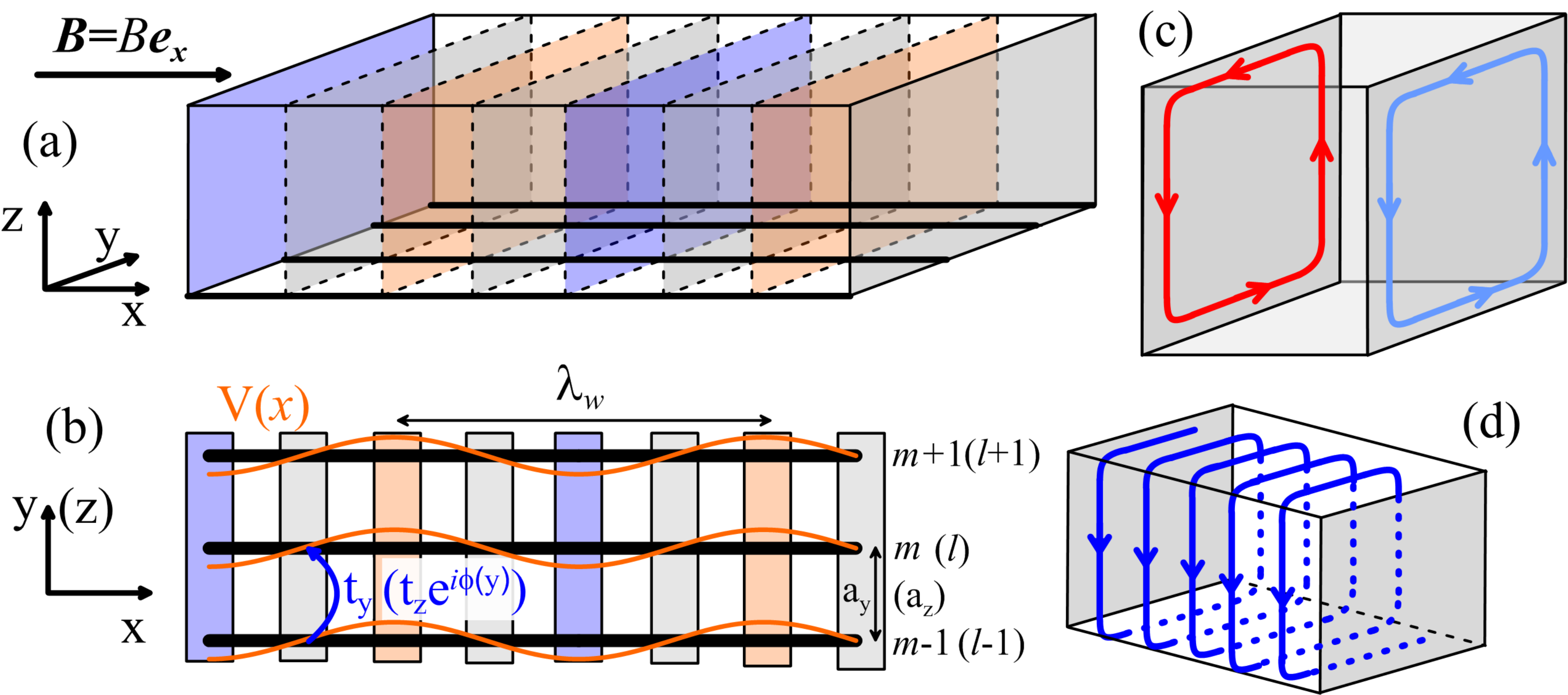}
\caption{ 	(a) A 3D stack of weakly coupled wires in a magnetic field $B $ applied along their axis.  (b) The wires (thick black lines) aligned along the $x$ axis and labeled by the indexes $m$ ($l$) are weakly coupled with tunnel amplitude $t_y$, $(t_z)$ in the $y$ ($z$) direction. The vector potential ${\bf A}=B y \boldsymbol{e}_z$ is chosen to be in the $z$ direction such that the corresponding tunneling phase $\phi (y)=eB a_z y / \hbar c $ is position-dependent, where $a_z$ is the distance between neighboring wires in the $z$ direction. In addition, there is a CDW modulation $V(x)$ [orange wavy line (b)] of the chemical potential of the wires with period $\lambda_w$.  The sketch (c) of the hinge modes  localized on the left (red) and right (blue) surfaces and (d)  of surface states localized on all surfaces except left and right.}
\label{setup}
\end{figure}

{\it Model.}
We consider a 3D  coupled-wire construction~\cite{lebed1986anisotropy, PhysRevB.43.11353, PhysRevLett.111.196401, PhysRevB.89.085101, PhysRevB.90.115426, PhysRevB.89.104523, PhysRevB.90.201102, PhysRevB.90.235425, PhysRevB.91.085426, PhysRevB.91.241106,SzumniakQH, ManishaQH, Meng_recent} 
in the presence of a uniform magnetic field applied along the wire axis, see Fig.~\ref{setup}. 
In addition, we include a CDW modulation along the wires [Fig.~\ref{setup}~(b)]. Such CDWs may be induced intrinsically by electron-electron interactions~\cite{CDW_interactions_1, CDW_interactions_2, CDW_interactions_3} or by an internal superlattice structure~\cite{CDW_superlattice_1,CDW_superlattice_2}, or extrinsically by periodically arranged gates inducing spatial modulations of the chemical potential~ \cite{CDW_luka}. The system is then described by the following tight-binding Hamiltonian,
\begin{align}
&H_{3D}=\sum_{n,m,l}\Big[ -t_x c^{\dag}_{n+1,m,l}c_{n,m,l} -t_y c^{\dag}_{n,m+1,l}c_{n,m,l}\\
&-t_z e^{im\phi}  c^{\dag}_{n,m,l+1}c_{n,m,l}- \frac{1}{2}(V(n)+\mu) c^{\dag}_{n,m,l}c_{n,m,l} + \text{H.c.}\Big]  \nonumber,
%\label{eq:H_2D}
\end{align}
where $c_{n,m,l}$ is the annihilation operator acting on the electron at a site $(n,m,l)$ of the lattice with the lattice constants $a_x$,  $a_y$,  $a_z$, respectively in the  $x$, $y$, $z$ directions. Here, without loss of generality, the hopping matrix elements $t_{x}$, $t_y$, $t_z$ are assumed to be real.
For simplicity, we consider spinless electrons in this work. A uniform magnetic field is applied in  $x$ direction, $\boldsymbol{B}=B\boldsymbol{e}_x$, and the corresponding vector potential, $\boldsymbol{A}=By\boldsymbol{e}_z$, is chosen along the $z$ axis, yielding the orbital Peierls phase $\phi=eBa_ya_z/\hbar c$.  The chemical potential is modulated in the presence of the CDW as $V(n)=2U_0\cos(2k_{w}na_x+\varphi)$ with the CDW amplitude $2U_0>0$ and the period $\lambda_{w}=\pi/k_{w}$. The angle $\varphi$ is the phase of the CDW at the left end of the wire ($n=0$).

With this choice of the vector potential $\bf A$, the system is translation-invariant in  the $z$ direction, thus, we can introduce the momentum $k_z$ via Fourier transformation $c_{n,m,l}=\frac{1}{\sqrt{N_z}}\sum_{k_z}c_{n,m,k_z}e^{-ilk_z a_z}$, where $N_z$ is the number of lattice sites in the $z$ direction. The Hamiltonian becomes diagonal in the $k_z$ space,
\begin{align}
&H(k_z)=\sum_{n,m,k_z}\Big ( [-t_x c^{\dag}_{n+1,m,k_z} c_{n,m,k_z}-t_y  c^{\dag}_{n,m+1,k_z} c_{n,m,k_z}\nonumber \\
&+\text{H.c}. ] - c^{\dag}_{n,m,k_z}c_{n,m,k_z} [\mu + V(n) + 2 t_z \cos(m\phi+k_za_z)] \Big ).
\label{eq:H_1D_ky}
\end{align}
As a result, the eigenfunctions of $H$ factorize as $e^{i k_z z }\psi_{k_z}(x,y)$, with $x=na_x$, $y=ma_y$, and $z=la_z$. From now on, we focus on $\psi_{k_z}(x,y)$ and treat $k_z$ as a parameter.

\begin{figure}[t!]
	\centering
	\includegraphics[width=8.6cm]{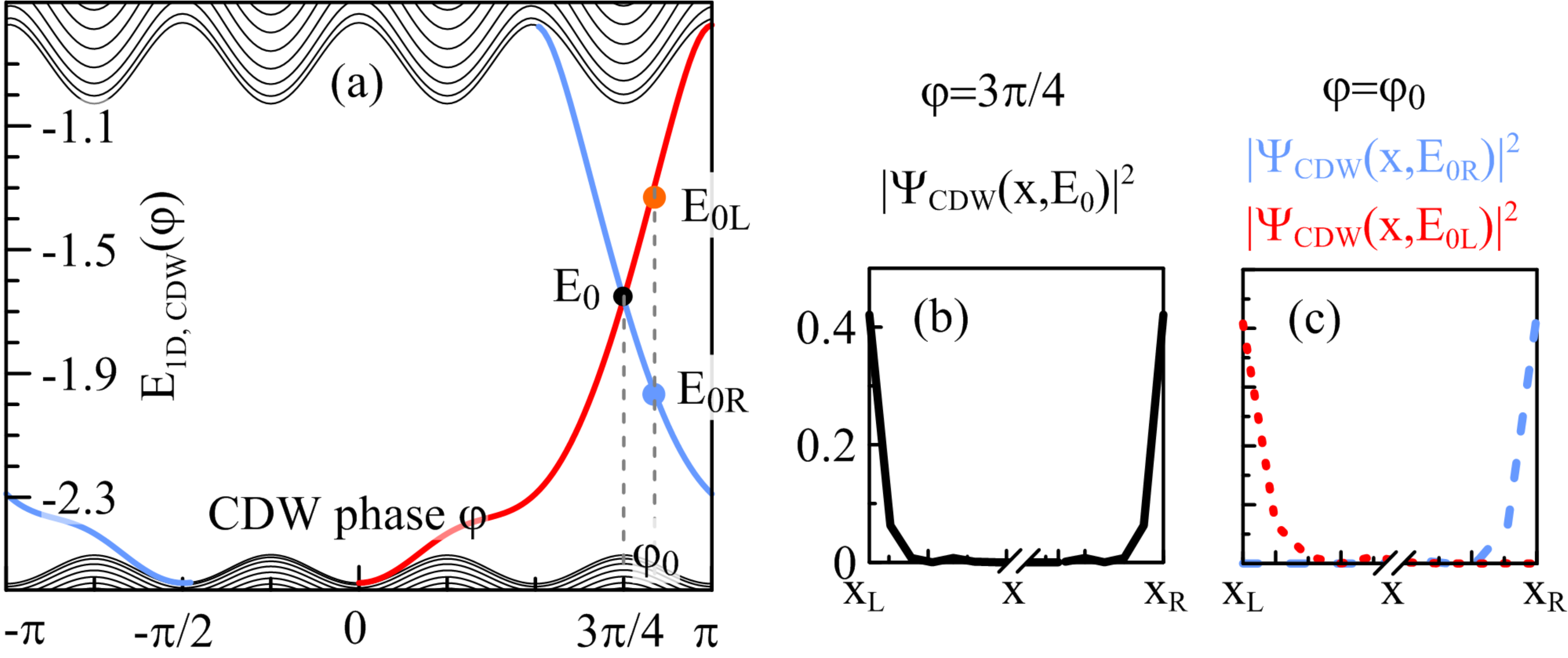}
	\caption{(a) Energy spectrum of a single wire with CDW modulation as function of phase $\varphi$. For $\varphi=3\pi/4$ the gap contains doubly degenerate end states at energy $E_0$ and for $\varphi=\varphi_0$ two non-degenerate end states at  energies $E_{0L}$ and $E_{0R}$. (b,c)~The corresponding probability densities  $|\Psi_{\text{CDW}}(x;E)|^2$ for $E=E_0, E_{0L}, E_{0R}$  indicate that double-degenerate states at  $E_0$  are localized at both ends of the wire, while states at energies $E_{0L}$ and $E_{0R}$ are localized at the left ($x_L$) and the right ($x_R$) end of the wire. Here, we choose $U_0=0.9t_x$ and $2k_wa_x=\pi/2$.}
	\label{Fig_CDW}
\end{figure}

{\it Ingredients.}
 The presence of a CDW along a single wire leads to an opening of a gap in the energy spectrum and for certain values of $\varphi$ to the emergence of in-gap end states (see Fig.~\ref{Fig_CDW}). We choose $2k_wa_x=\pi/2$ for all plots to follow. For this choice, the system is gapped  at the  filling fractions $1/4$ and $3/4$. In Fig.~\ref{Fig_CDW}(a), we plot a part of the spectrum around the lowest gap as a function of $\varphi$. At the special values of $\varphi=-\pi/4$, $ 3\pi/4$, the system has an inversion symmetry. For $\varphi=-\pi/4$, there are no states in the lowest energy gap, while for $\varphi=3\pi/4$  two degenerate end states [localized at the both ends of the wire, see Fig.~\ref{Fig_CDW}(b)] emerge in the gap at energy $E_0$ \cite{CDW_luka, CDW_park,CDW_inversion_symmetry}. If one tunes away from $\varphi=3\pi/4$, the in-gap states split in energy.
  The states with $\partial E(\varphi)/\partial \varphi>0$ ($\partial E(\varphi)/\partial \varphi<0$) are localized at the left (right) end of the wire at  $x_L$ ($x_R$) as illustrated in Fig.~\ref{Fig_CDW}(c). At other values of $\varphi$, the system is fully gapped and there are no bound states.

\begin{figure}[bt!]
	\centering
	\includegraphics[width=8.6cm]{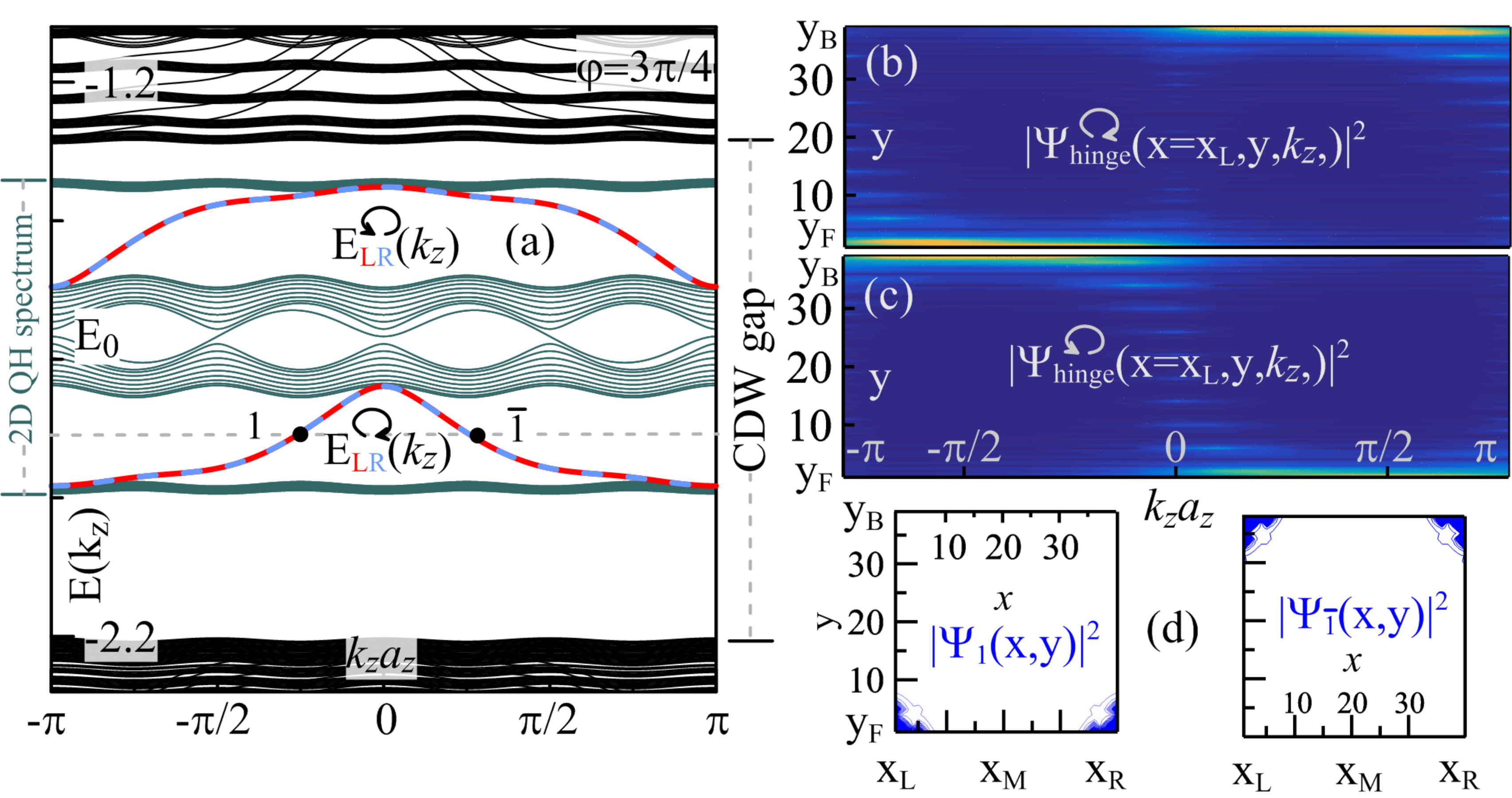}	
	\caption{(a) The  spectrum of the system from Fig.~\ref{setup}(a) with periodic boundary conditions in the $z$ direction around the lowest CDW-induced gap for $\varphi=3\pi/4$.  Two degenerate copies of a 2D QHE spectrum are centered around $E_0$ within two QHE gaps. The 2D QHE bulk bands (green) and 1D hinge modes $E_{LR}^{\sigma}$ (dashed red-blue) are localized at the left and right surfaces  with $\sigma=\circlearrowright (1)$ and $\sigma= \circlearrowleft (-1)$ denoting the propagation direction  of the hinge modes.
(b,c)	The color maps  represent the cross sections of the QHE hinge modes  probability density $|\Psi_{\text{hinge}}^{\sigma}(x=x_L,y,k_z)|^2=|\Psi_{\text{hinge}}^{\sigma}(x=x_R,y,k_z)|^2$. (d) The probability density $|\Psi_{1}(x,y)|^2$, $|\Psi_{\bar{1}}(x,y)|^2$ for the QHE hinge modes at selected $k_z$ points, $1$ and $\bar{1}$,  indicated in (a). Here, we set $t_x=1$, $t_y=0.11t_x$, $t_z=0.09t_x$, $U_0=0.9t_x$, $2k_wa_x=\pi/2$ with $N_x \times N_y=40\times39$ lattice points and take a resonant magnetic field $\phi=2k_Fa_y=\pi/2$, corresponding to filling factor $\nu=1$. }
	\label{Fig3pi_4}
\end{figure}

On the other hand,  electrons in a single 2D $yz$ layer and in the presence of a perpendicular magnetic field exhibit the well-known Hofstadter  spectrum~\cite{Hofstadter} with chiral edge states. 
For illustrative purposes, we consider the simplest case with a resonant magnetic field value leading to $\phi =\pi/2$, corresponding to the QHE filling factor $\nu = 1$. (However, we emphasize that our results are valid for  other values of magnetic flux with higher QHE filling factors and for  other periods of the CDW potential as we show in the Supplementary Material~(SM)~\cite{SM}).
In our proposal, we combine these two mechanisms, which leads to opening of  gaps and emergence of  in-gap states that are exponentially localized on the hinges of the left ($x=x_L$,$y$,$z$) [$(\bar{1},0,0)$] and right ($x=x_R$, $y$,$ z$) [$(0,0,1)$] surfaces, see Fig. \ref{Fig3pi_4}. 
For the choice $2k_wa_x=\pi/2$, the spectrum of the  3D system is characterized by two CDW-induced gaps (at 1/4 and 3/4 filling). We take the amplitude of the CDW potential $U_0$ to be sufficiently large so that the 2D QHE spectrum (or at least one of the QHE gaps)  fully fits inside the CDW-induced gap.  This requires $U_0\gg4t_y$. The effects of different values of $U_0$ on the spectrum is discussed in the SM~\cite{SM}.\\

\begin{figure}[bt!]
	\centering
	\includegraphics[width=8.6cm]{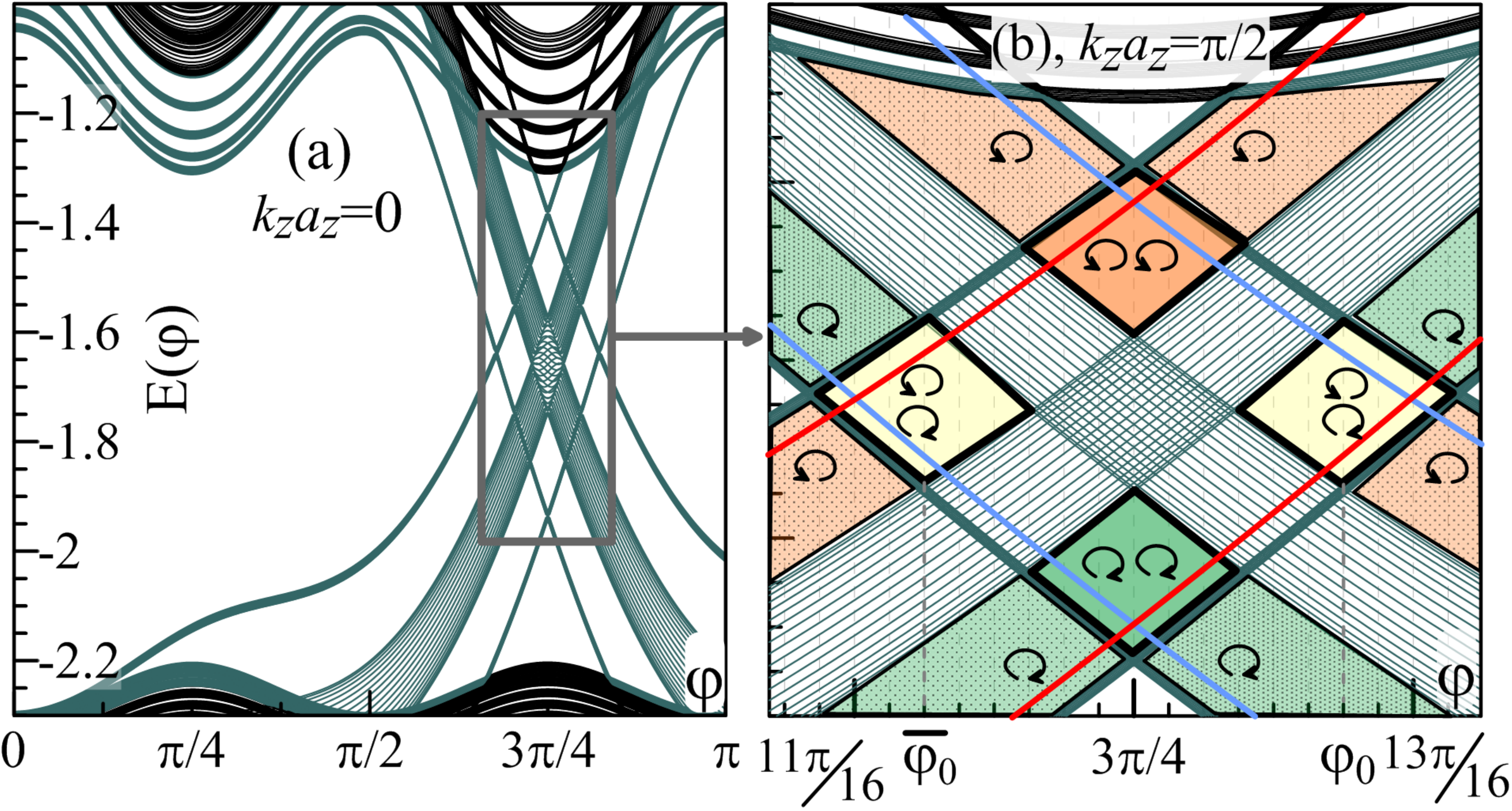}
	\caption{(a) Spectrum of the 3D system from Fig.~\ref{setup}(a) around the CDW-induced gap as a function of $\varphi$ for $k_z=0$. The interesting part, where the QHE spectra of states localized at the left and right surfaces cross and overlap, is marked by a gray rectangle and zoomed in on panel (b) for $k_z a_z=\pi/2$. The pale orange (green) color denotes QHE gaps with a single hinge mode characterized by counterclockwise $\sigma=1=\circlearrowleft$ (clockwise $\sigma=-1=\circlearrowright $) direction of propagation. The dark orange (green) areas denotes QHE gaps with pair of chiral unidirectional QHE hinge modes with $\sigma=1$ ($\sigma=-1$) localized at the left and right surfaces. The yellow area denotes a scenario with a pair of in-gap hinge modes propagating in opposite directions on the left and right surface. The solid red (blue) lines correspond to hinge modes localized at hinges adjacent to the left (right) surface. }
	\label{phase_diagram}
\end{figure}

{\it Degenerate QHE hinge modes.}
First we consider the case with $\varphi=3\pi/4$ for which a single CDW modulated wire has a spectrum with two degenerate in-gap end states at energy $E_0$. In this case, we find  doubly degenerate copies of the 2D QHE spectrum inside the CDW-induced gap centered around $E_0$  [see Fig.~\ref{Fig3pi_4}~(a)]. 
The quasi-2D QHE bulk states [green lines in Fig.~\ref{Fig3pi_4}~(a)] are exponentially localized at the right and left surfaces of the system. One can see two QHE gaps ($\nu=1$) with doubly degenerate hinge  modes $E_{LR}^{\sigma}(k_z)$, where the index $\sigma=\circlearrowright, \circlearrowleft\equiv-1,1$ corresponds to the direction of propagation/circulation in the $yz$-plane for open boundary conditions in both the $y$ and $z$ directions.
The hinge modes are chiral as indicated by the probability densities $|\Psi_{\text{hinge}}^{\sigma}(x=x_L,y,k_z)|^2=|\Psi_{\text{hinge}}^{\sigma}(x=x_R,y,k_z)|^2$ [see Fig.~\ref{Fig3pi_4}~(b,c)] and are exponentially localized at certain corners of the $xy$-cross section [see Fig.~\ref{Fig3pi_4}(d)]. For a finite-size system with open boundary conditions in all directions, the  hinge modes circulate in the same direction on the hinges adjacent to the left and right surface of the system [as schematically illustrated in Fig.~1(c)], in analogy to unidirectional edge states in 2D topological systems~\cite{unidi_1, unidi_2, unidi_3}.

\begin{figure}[t!]
	\centering
	\includegraphics[width=8.6cm]{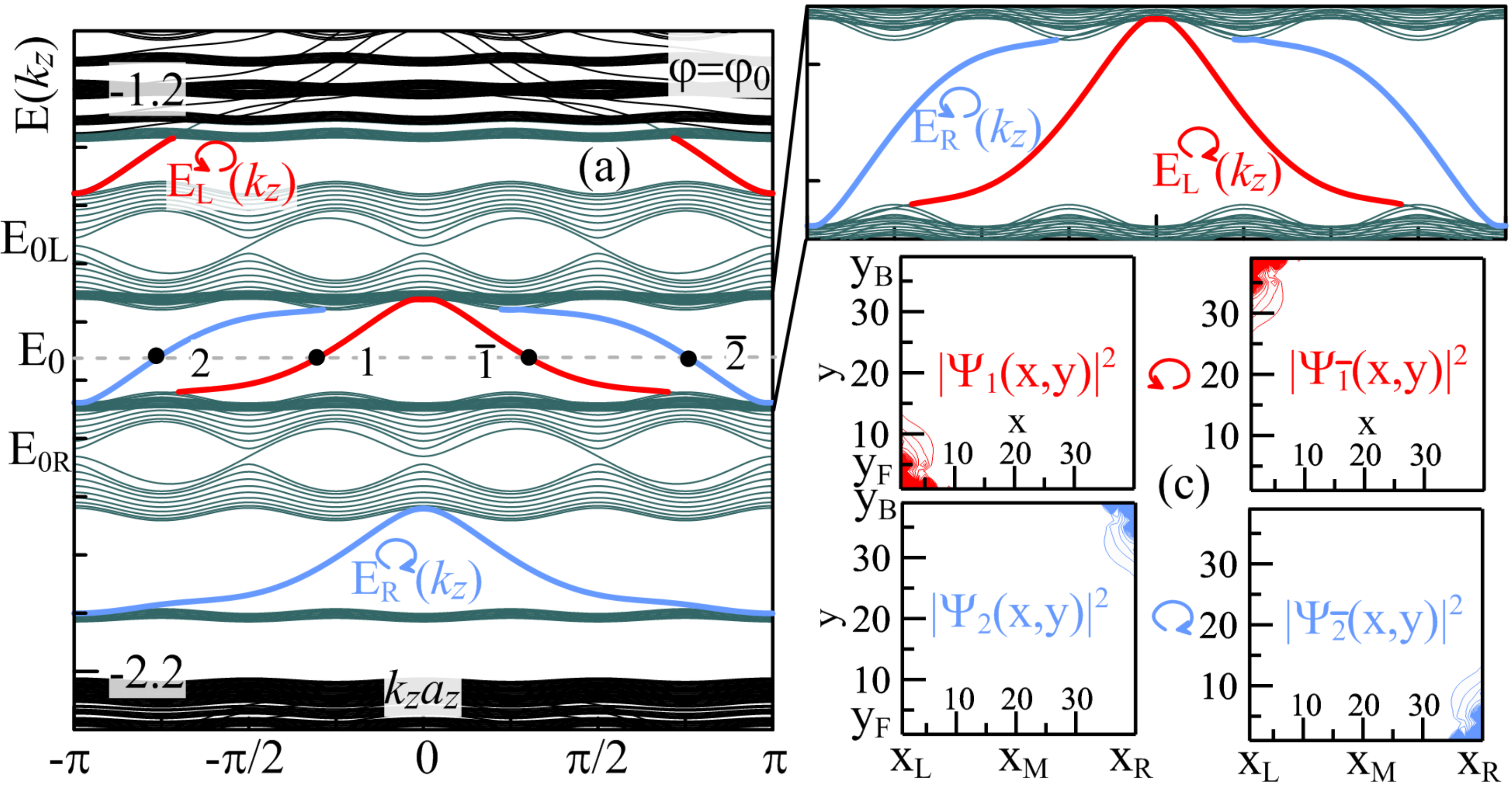}	
	\caption{Same as in Fig. \ref{Fig3pi_4} but for $\varphi=\varphi_0$. The two degenerate copies of the bulk QHE spectrum from Fig. \ref{Fig3pi_4} are split in energies such that each one is centered around  $E_{0R}$ and $E_{0L}$,  respectively. The part of the QHE spectrum around  $E_{0R}$  ($E_{0L}$)  corresponds to states localized on left (right) surface. Interestingly, for the chosen CDW phase $\varphi=\varphi_0$, in the gap around $E_0$ we find a pair of chiral hinge modes that propagate in the opposite directions on the left and the right surface.   (c) The corresponding wave function probability $\Psi_{1,\bar{1}, 2, \bar{2}}$ for $k_z$ momenta marked by 1, $\bar{1}$, 2, $\bar{2}$ points on panel (a). }
	\label{Fig_fi0}
\end{figure}

{\it Hinge mode regimes.} 
In order to systematize the regimes in which hinge modes exist,
we plot the spectrum as  function of $\varphi$ for fixed $k_z=0$ or $\pi/2$ in Fig.~\ref{phase_diagram}. The QHE gaps marked by pale orange (green) host only one hinge mode with the counterclockwise  $\sigma=\circlearrowleft$ (clockwise $\sigma=\circlearrowright$) direction of propagation. The dark orange (green) regions around $\varphi=3\pi/4$ correspond to a scenario where QHE gaps of each QHE copy overlap thus hosts two hinge modes (one at the left  and one at the right surface) with the same positive (negative) $\sigma$. For $\varphi=3\pi/4$, the QHE gaps of each QHE  copy fully overlap.
If one  deviates from  $\varphi=3\pi/4$, the degeneracy of the QHE  spectrum is lifted. Each of the 2D QHE copies is now shifted in energy, either up or down and corresponding states are localized at the left or right surfaces, consistent with the picture emerging from 1D CDW-induced end states in Fig.~\ref{Fig_CDW}. The hinge modes are circulating again in the same directions on each surface (see the SM~\cite{SM}). Moreover, further increase of $\varphi$ leads to closing of the QHE gaps in such a way that the gap of  one of the QHE copies overlaps  in energy with the 2D QHE bulk states of the other copy localized at the opposite surface. 

Interestingly, there is a region around $\varphi=\varphi_0$ and $\bar{\varphi_0}$ (marked by  yellow rectangles), where the hinge modes of opposite  $\sigma$ coexist (see Fig.~\ref{Fig_fi0}).   The system hosts hinge modes  propagating in  opposite directions on  hinges adjacent to the left and right surface. The direction of propagation can be reversed by changing $\varphi$ from a region around $\varphi_0$ to one close to $\bar{\varphi_0}$ without the need of changing $\mu$. The overlap of the QHE gaps is maximal for $\varphi=\varphi_0$ and $\bar{\varphi_0}$. The edge states localized on the hinges adjacent to the left (right) surfaces are marked by red (blue) solid lines. Further increase of $\varphi$ away from $\varphi=3\pi/4$ leads to the separation in energy of these two  QHE copies.
We also checked numerically that hinge modes are robust against moderate static disorder (as  shown in the SM~\cite{SM}), however, we note that for the special case of $\varphi=3\pi/4$ the disorder lifts the degeneracy of the hinge modes on the left and right surfaces.

\begin{figure}[bt!]
	\centering
	\includegraphics[width=8.6cm]{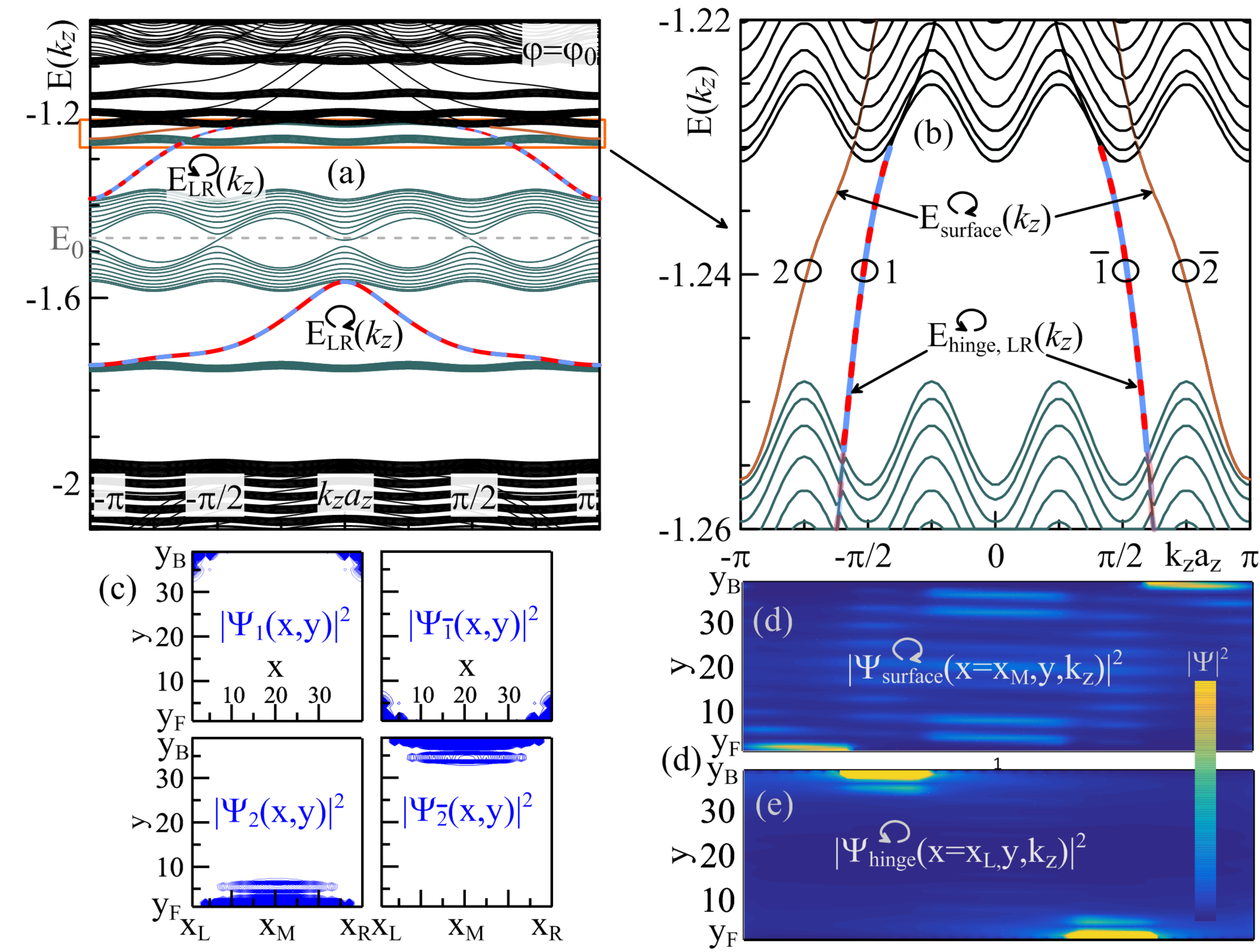}
	\caption{ (a) Same as in Fig. \ref{Fig3pi_4}(a) but for smaller $U_0=0.7t$.  (b)  The zoomed-in part of the spectrum [marked by  orange  rectangle on (a)] contains the gap in which doubly degenerate  chiral QHE hinge modes $\text{E}_\text{hinge}(k_z)$ and nondegenerate QHE  chiral surface  states $\text{E}_\text{surface}(k_z)$ coexist. (c) The corresponding probability density $|\Psi_{\text{surface}}(x,y,k_z^{\text{fixed}})|^2$ for fixed $k_z$ denoted by $1, \bar{1}, 2, \bar{2}$.  (d,e) We also plot the probability density  as a function of $y$ and $k_z$, $|\Psi_{\text{surface}}(x=x_\text{M},y,k_z)|^2$ and $|\Psi_{\text{hinge}}(x=x_L,x_R;y;k_z)|^2$, to demonstrate the chiral character of the states. 
	}
	\label{hybrid}
\end{figure}

{\it Hybrid scenario.}
Another striking feature of the proposed setup is the fact that it can support a hybrid scenario with mixed high-order topology~\cite{Bultinck_hybrid_topology}, in which the 2D QHE chiral surface states can coexist in the gap with the 1D QHE hinge modes. Such a case can be realized if we allow for the hybridization of the quasi-2D QHE states with the  bulk bands of higher energy. This can be achieved by pushing the 3D bulk spectrum into 2D QHE spectrum by reducing $U_0$ or vice versa by pushing the 2D QHE spectrum into 3D bulk spectrum by increasing $t_y$.  We find that reducing $U_0$ is a more efficient way to reach the hybrid phase as we demonstrate in the SM \cite{SM}. In Fig.~\ref{hybrid}, we show a part of the spectrum of the 3D system in which one can see a pair of the chiral QHE surface states localized at the  front~$(x,y_F, z)$ and back~$(x,y_B, z)$ surfaces and propagating in the $z$ direction. [For the open boundary conditions along  the $z$ axis, the 2D QHE chiral surface states are circulating on  all surfaces except the left and right one {as sketched in Fig.~1(d).] Interestingly, we can also find coexisting pairs of doubly degenerate chiral QHE hinge modes localized at the  corners of the finite $xy$ plane and propagating  in the $z$ direction. Since the hinge modes propagate in opposite direction to the surface states one can expect backscattering at the corners which can lead to local reduction of the conductance.

{\it Summary.} 
We have studied a system composed of a stack of coupled CDW modulated wires in ther presence of an external magnetic field oriented parallel to the wire axis. We showed that such a system can support a second-order topological QHE phase with hinge modes localized on the surfaces perpendicular to the applied magnetic field. Quite remarkably, the direction of propagation of the hinge modes can be switched by tuning the phase of the CDW without the need of reversing the direction of the applied magnetic field. The hinge modes are immune to moderate static disorder~\cite{SM}.
Furthermore we demonstrated that the proposed system can support hybrid higher-order topology with QHE surface and hinge modes coexisting in the gap.
We propose that our predictions can be tested in semiconducting nanowires with CDW modulations, heterostructures forming a superlattice of 2DEGs with electrically tunable  chemical potentials~\cite{sup_exp_1,sup_exp_2}, organic conductors \cite{org_exp}, and optical lattices~\cite{optlatt_exp_1, optlatt_exp_2, optlatt_exp_3} or photonic crystals~\cite{pcryst_exp}. Finally, we expect that the CDW mechanism uncovered here will also give rise to hinge modes in other 3D topological systems composed of stacked CDW-modulated layers.

\begin{acknowledgments}
	
We acknowledge support within HOMING programme of the FNP co-financed by the European Union under the European Regional Development Fund, Swiss National Science Foundation, and NCCR QSIT. This project received funding from the European Unions Horizon 2020 research and innovation program (ERC Starting Grant, grant agreement No 757725). The calculations were performed on PL-Grid Infrastructure.
\end{acknowledgments}
\bibliographystyle{unsrt}

\onecolumngrid

%\newpage
\vspace*{1cm}
\begin{center}
	\large{\bf Supplemental Material for ``Hinge Modes and Surface States in Second-Order Topological Three-Dimensional Quantum Hall Systems induced by Charge Density Modulation  '' \\}
\end{center}
\begin{center}
	Pawe\l{} Szumniak,$^{1}$ Daniel Loss,$^2$ and Jelena Klinovaja$^2$\\
	{\it $^1$ AGH University of Science and Technology, Faculty of
	Physics and Applied Computer Science,}\\
	{\it al. Mickiewicza 30, 30-059 Krak\'ow, Poland}
	{\it $^2$Department of Physics, University of Basel, Klingelbergstrasse 82, CH-4056 Basel, Switzerland}
\end{center}
%\vspace*{1cm}
\onecolumngrid
\setcounter{equation}{0}
\setcounter{figure}{0}
\section{Chiral surface states in 3D CDW modulated QHE system}\label{surface}
In this section, we focus on the special case with $\varphi=-\pi/4$. For this choice of parameter,  there are no states in the CDW gap. However, interestingly, we can notice [see Fig.~\ref{Figmpi_4}~(b)] that the part of the spectrum around the upper edge of the CDW-induced gap (marked with the orange rectangular) contains gaps that host chiral 2D QHE surface states. This happens also for other values of  $\varphi$. Such 2D QHE surface states are propagating in "$+z$" and "$-z$" direction and are localized respectively on the front ($x,y=y_F, z$) and back ($x,y=y_B, z$) surfaces [see Fig.~\ref{Figmpi_4}(e)]. We observe a series of small gaps in which one can find from one up to four 2D QHE chiral surface modes. This gaps are opened due to the interplay of the CDW and the magnetic field. The size of these gaps does not depend on the CDW phase $\varphi$, however, their position does. In the area between the gaps, the surface QHE modes hybridize with the 3D bulk states.  From Figs.~\ref{Figmpi_4}(c)-(d), one can conclude that surface states are chiral, i.e. for the case of  periodic boundary conditions in $z$ direction, the 2D QHE states with $\partial E(k_z)/\partial k_z>0$ ($\partial E(k_z)/\partial k_z<0$) are localized at the front (back) surface.\\

\begin{figure*}[b!]
	\centering
	\includegraphics[width=17cm]{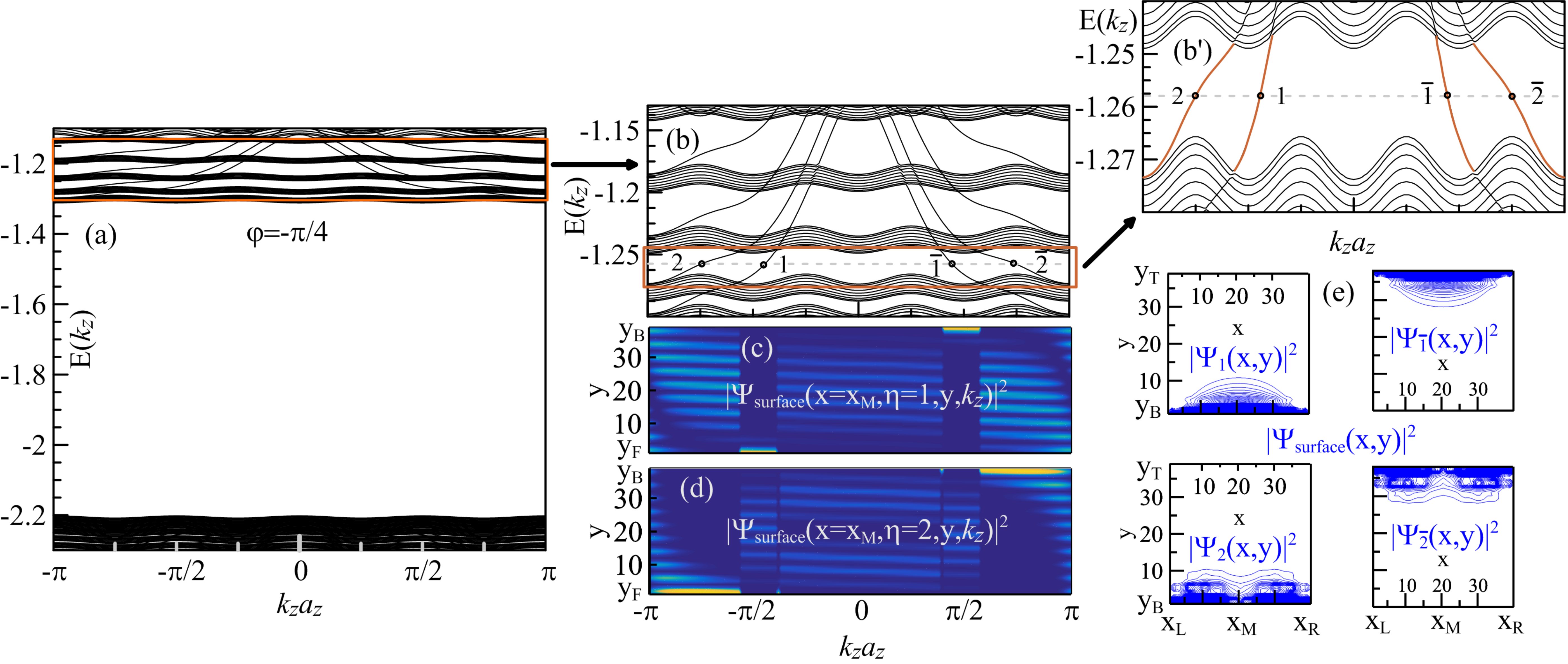}	
	\caption{ (a) Spectrum of the 3D system from Fig.~1~(a) from the main text with the periodic boundary conditions in the $z$ direction around the lowest in energy CDW induced gap as a function of $k_z$ for $\varphi=-\pi/4$. The zoom of upper edge of CDW gap is plotted in the panel (b) and further zoom of one of the gaps in (b'). The color maps (c,d) shows probability density $|\Psi_{\text{surface}}(x=x_M,y, k_z)|^2$  for the first and second QHE surface mode hosted in the gap marked by the brown rectangle and further zoomed in (b'). The probability density $|\Psi_{\text{surface}}(x,y, k_z^\text{fixed})|^2$  for the first ($\eta=1$) and second mode ($\eta=2$) of the QHE chiral surface states [selected $k_z^\text{fixed}$ points: $1$, $\bar{1}$, $2$, $\bar{2}$ from panels (b), (b')  is represented on panel (e)]. Here, we set $t_y=0.11$, $t_z=0.09$, $U_0=0.9t_x$, $2k_wa_x=\pi/2$ with $N_x \times N_y=40\times39$ lattice points and take the resonant value for the magnetic field $\phi=\pi/2$.}
	\label{Figmpi_4}
\end{figure*}

\newpage
\section{Stability}
Here, we study the stability of the hinge modes in the presence of disorder. We also demonstrate that the hinge modes are present in the spectrum for a wide range of  system parameters.

\subsection{Disorder}\label{Disorder}
\begin{figure}[ht!]
	\centering
	\includegraphics[width=10cm]{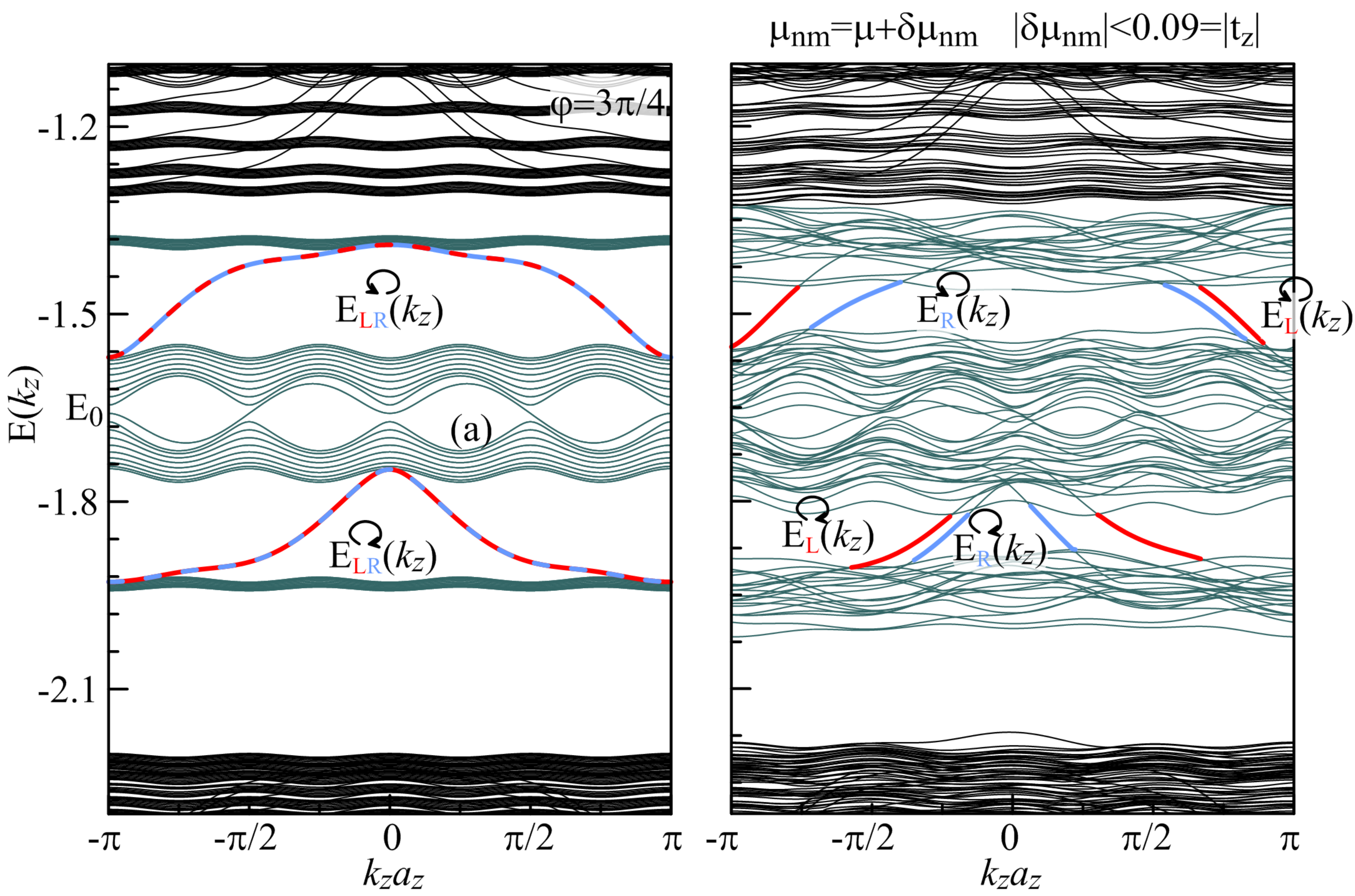}
	\caption{Spectrum of the 3D system from Fig.~1(a) of the main text with  periodic boundary conditions in  the $z$ direction around the energetically lowest  CDW-induced gap as a function of $k_z$ for $\varphi=3\pi/4$: (a) without disorder and (b) in the presence of onsite static disorder corresponding to a uniformly fluctuating chemical potential in the range of the QHE gap $|\delta \mu_{n,m}|\leq0.09=|t_z|$. Despite the fact that the bulk gap gets smaller, the hinge modes are clearly present in the spectrum.}
	\label{Fig3pi_4_disorder}
\end{figure}

First, we investigate the effects of disorder on the stability of the hinge modes. Again, we consider static disorder - random onsite fluctuations in the chemical potential of the order of the QHE gap size: $|\delta \mu_{n,m}|\leq0.09=|t_z|$. In the presented case, for computational purposes,  disorder is translationally invariant in the $z$ direction, $\delta \mu_{n,m}(k_z)=\delta\mu_{n,m}$. One can see [Fig.~\ref{Fig3pi_4_disorder}] that bulk bands are strongly affected and the QHE gaps are reduced with respect to the clean system, however, there is  still a region of $k_z$  values for which hinge modes exist in the gap. One can also notice that disorder lifts the accidental degeneracy between left and right surfaces. Interestingly, the hinge modes are quite stable even though  they exist in  gaps much smaller than the fluctuations of the chemical potential.

\newpage
\subsection{Chiral hinge modes: lifting degeneracy between right and left surfaces}
Here, we show an example of a regime where there are two non-degenerate hinge modes in the gap. This scenario is realized for $\varphi$ around $\varphi=3\pi/4$, i.e. $\varphi=3\pi/4+\delta\varphi$ (in the region marked by the dark green and orange in Fig.~4(b) of the main text). In this case, the degeneracy is lifted, the QHE gap is reduced, and hinge modes that are localized on the left and the right surfaces are split in energy. Furthermore, there are two more gaps that emerge in the spectrum which host a single QHE hinge mode on one of the surfaces [see pale orange and green areas in Fig.~4(b) of  the main text]. 

\begin{figure}[ht!]
	\centering
	\includegraphics[width=8.6cm]{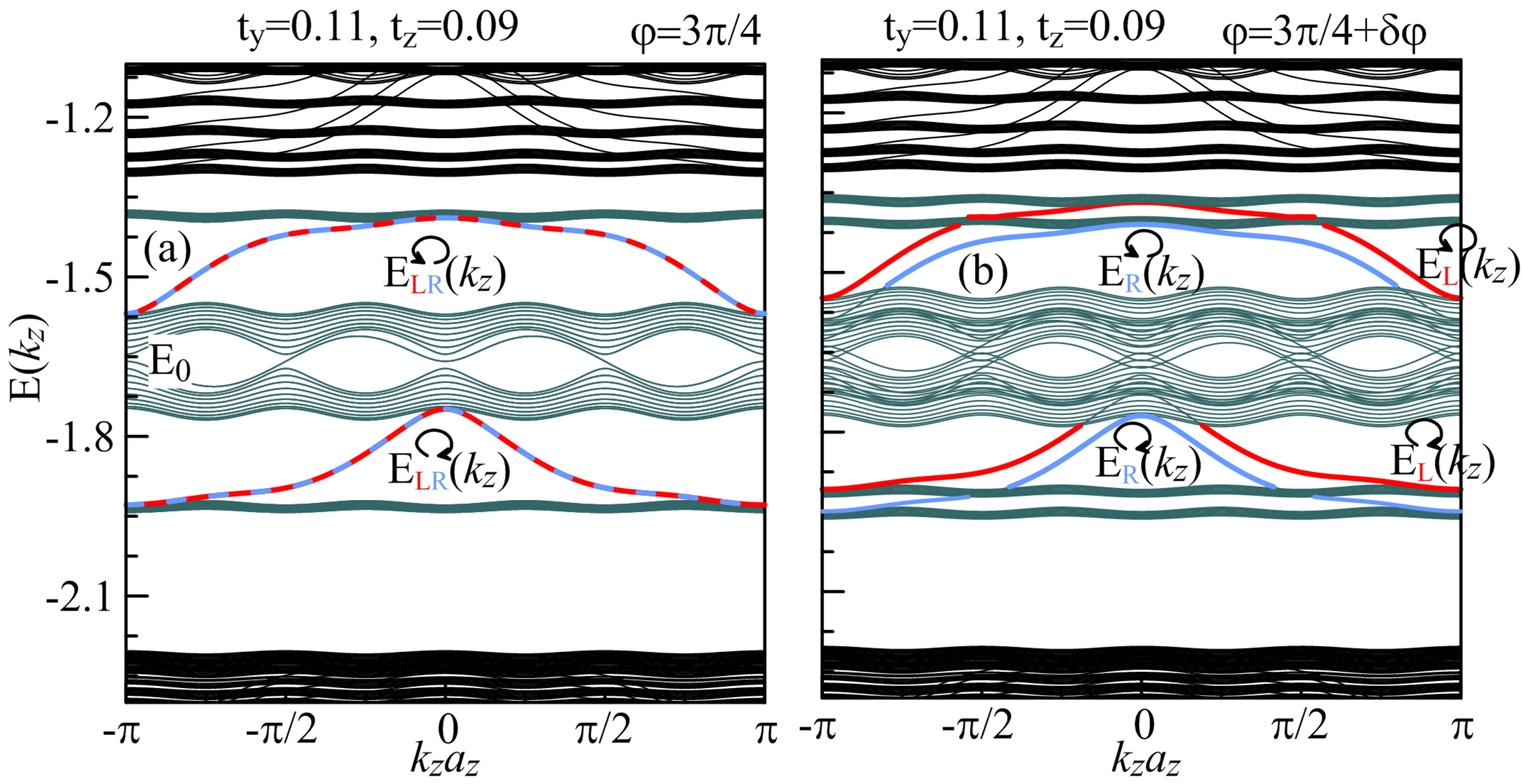}
	\caption{Spectrum of the 3D system from Fig.~1(a) of the main text with  periodic boundary conditions in  the $z$ direction around the energetically lowest CDW-induced gap as a function of $k_z$ for (a) $\varphi=3\pi/4$ and (b) $\varphi=3\pi/4+\delta\varphi$. One can clearly see in panel (b) that by moving away from $\varphi=3\pi/4$, e.g. for $\varphi=3\pi/4+\delta\varphi$,  the degeneracy of the two QHE spectrum copies is lifted, which results in splitting in energy of the hinge modes and bulk bands belonging to two different surfaces. Now, the QHE gaps are reduced and host two hinge modes localized on the left (red curve) and right (blue curve) surfaces which corresponds to the regimes marked by dark green and orange on Fig.~4(b) of the main text. 
	}
	\label{Fig3pi_4_dfi}
\end{figure}

\subsection{Spectrum dependence on $U_0$ and $t_y$}

\begin{figure*}[ht!]
	\centering
	\includegraphics[width=14cm]{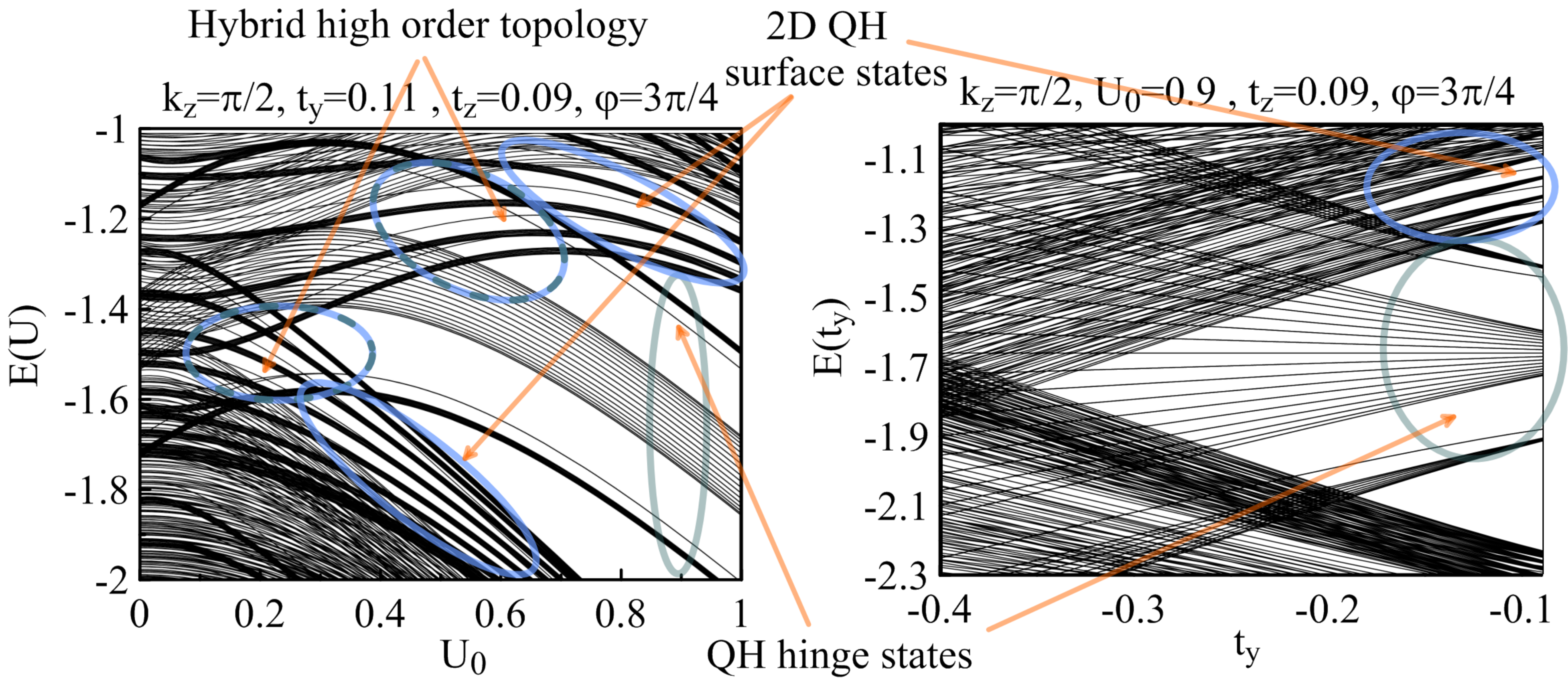}	
	\caption{Cross-section of the spectrum of the 3D system from Fig.~1(a) of the main text with periodic boundary conditions in  the $z$ direction as a function of (a) $U_0$ and (b) $t_y$ for fixed $\varphi=3\pi /4$, $k_z=\pi/2$,  $t_z=0.09$, and (a) $t_y=0.11$, (b) $U_0=0.9$.  The interesting regimes are marked schematically by ovals. For $U_0\gtrsim0.8$, the part of the spectrum with  hinge modes is well separated from one with QHE surface states. If  $U_0$ is decreased, the 2D QHE spectrum with the QHE gap hosting quasi-1D hinge modes is pushed into the 3D bulk states and starts to overlap with gaps hosting 2D QHE chiral surface states. In such a regime, the scenario with a hybrid topology is realized. Further decrease of $U_0$ below $U_0\lesssim 0.4$ leads to an emergence of an another region with mixed higher-order topology. On the other hand, when $|t_y|$ is increased, the QHE hinge modes start to hybridize with the QHE surface states. However, in this case, the bulk states mask them, making them undetectable in transport experiments.
	}
	\label{Fig_U_ty}
\end{figure*}

We can see that for $U_0\gtrsim0.8$ the part of the spectrum with QHE hinge states is well separated from QHE surface states. By decreasing $U_0$, one can see that the upper QHE gap with quasi-1D hinge states starts to overlap with gaps hosting 2D QHE surface states. In this case, the scenario with hybrid topology is realized. Further decrease of $U_0$ below $U_0\lesssim 0.4$ leads to the emergence of an another region with mixed high order topology.
In the left panel of Fig.~\ref{Fig_U_ty}, we explore the  spectrum as a function of $t_y$ with  other parameters being fixed: $\varphi=3\pi /4$, $k_z=\pi/2$,  $t_z=0.09$, and $U_0=0.9$. As  $|t_y|$ is increased, QHE hinge states starts to hybridize with the QHE surface states. However, in this case, the 3D bulk states mask both states, making them undetectable in transport experiments.

\subsection{Isotropic and anisotropic system in the $yz$ plane}\label{Anisotropic}
Here, we analyze effects of anisotropy in the hoping amplitude in the $yz$ plane. To be specific, we consider the case with $t_y=t_z=0.11$; $t_y=0.11$, $t_z=0.09$, and $t_y=t_z/2=0.055$.
We can notice that  decreasing $t_z$ with respect to $t_y$ causes flattening of the bulk bands. The despersion relation of the hinge modes in the strongly anisotropic limit has a cosine-like shape, which can be derived from the linearized model \cite{JKLinearized,Meng}.

\begin{figure*}[ht!]
	\centering
	\includegraphics[width=16cm]{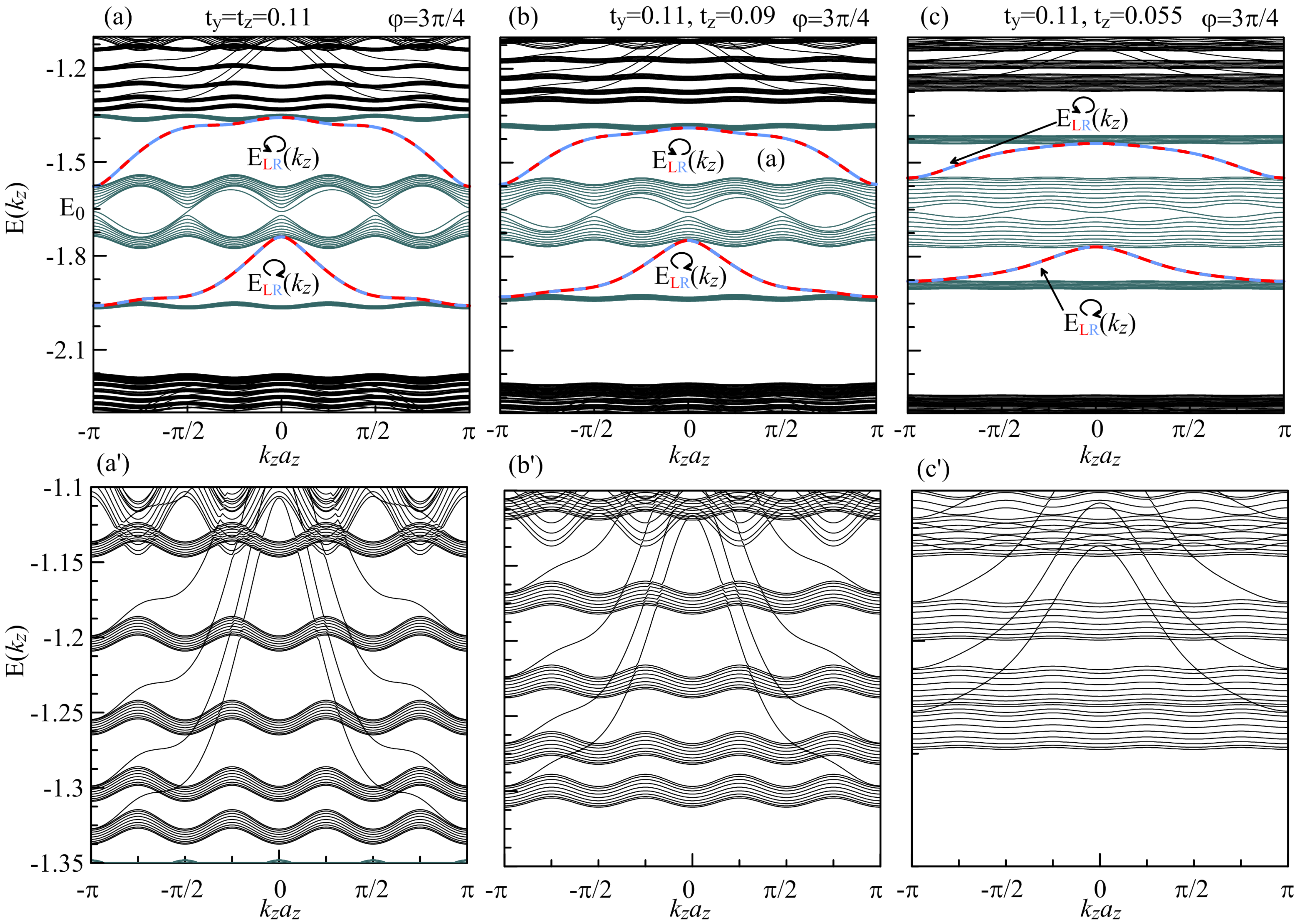}	
	\caption{Spectrum of the 3D system from Fig.~1(a) of the main text with  periodic boundary conditions in  the $z$ direction around the energetically lowest CDW-induced gap as a function of $k_z$ for $\varphi=3\pi/4$  (a) for  isotropic hopping in the $yz$ plane with $t_y=t_z=0.11$, (b) for a slightly anisotropic case like in the main text with $t_y=0.11$, $t_z=0.09$,  and (c) for a highly anisotropic case with $t_y=t_z/2=0.055$. The lower panels (a'-c') show the zoomed-in part of the spectrum with the 3D QHE gaps hosting nondegenerate 2D QHE chiral surface states. One can clearly see that decreasing  $t_z$ results in flattening of the bulk bands and in reducing the size of the gaps. Moreover, for $|t_y|\gg|t_z|$,  (c) the chiral QHE  hinge modes and (c') chiral  QHE surface states have a cosine-like shape.}
	\label{Fig3pi_4_tz_iso_aniso}
\end{figure*}

\newpage
\subsection{Different periodicity of the CDW: $k_w=\pi/5a_x$}
The single CDW-modulated wire supports end states  for  different values of the period.
Here, we consider an example with $2k_wa_x=2\pi/5$,  for which one obtains a pair of degenerate end states for the CDW phase equal to $\varphi=4\pi/5$, see Fig.~\ref{Fig3pi_4_kw_pi_5}(a-c). In this case, for the 3D CDW-modulated system shown in  Fig.~1 of the main text, one observes a similar behaviour of the hinge modes [see Fig.~\ref{Fig3pi_4_kw_pi_5}(d,e)] as obtained before in the main text for $2k_wa_x=\pi/2$ and $\varphi=3\pi/4$ [see Fig.~3~(a,d) of the main text].

\begin{figure*}[ht!]
	\centering
	\includegraphics[width=16cm]{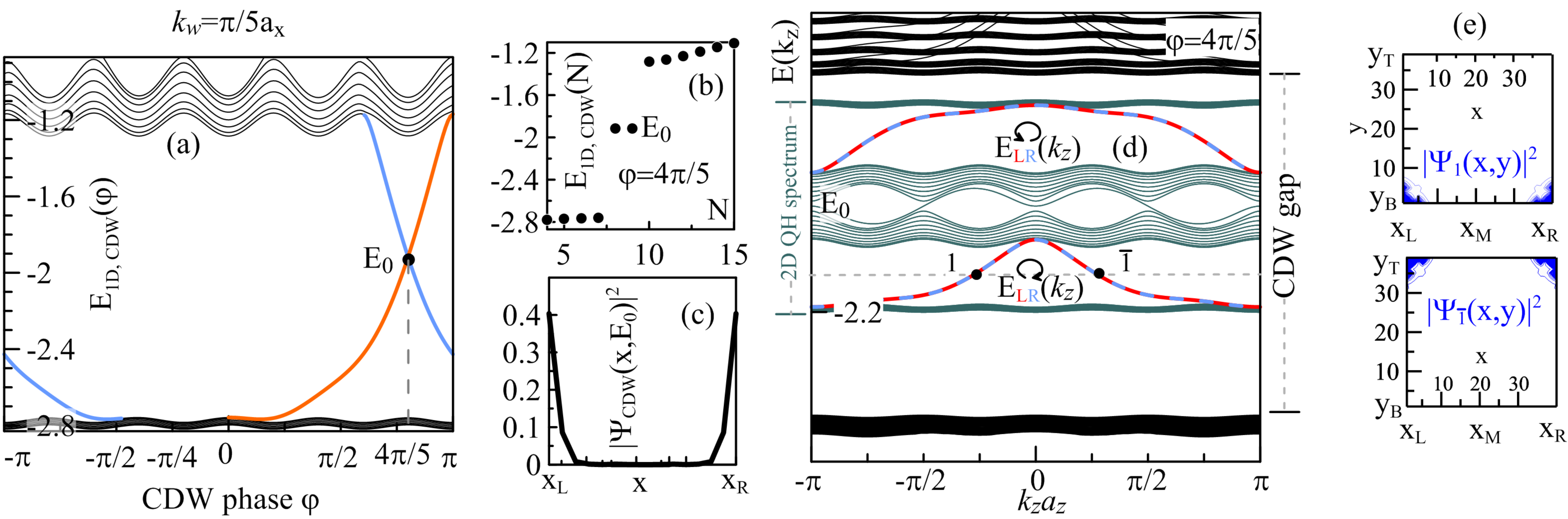}
	\caption{(a-d) The same as in Fig.~2(a, b, d) of the main text but for a different periodicity of the CDW potential, i.e. $k_w=\pi/5a_x$. In this case, the system has an inversion symmetry for $\varphi=4\pi/5$ and supports doubly degenerate hinge modes in the same way as we presented in Fig.~3(a,d) of the main text.}
	\label{Fig3pi_4_kw_pi_5}
\end{figure*}

\section{Higher QHE filling factors}

\begin{figure}[ht!]
	\centering
	\includegraphics[width=6cm]{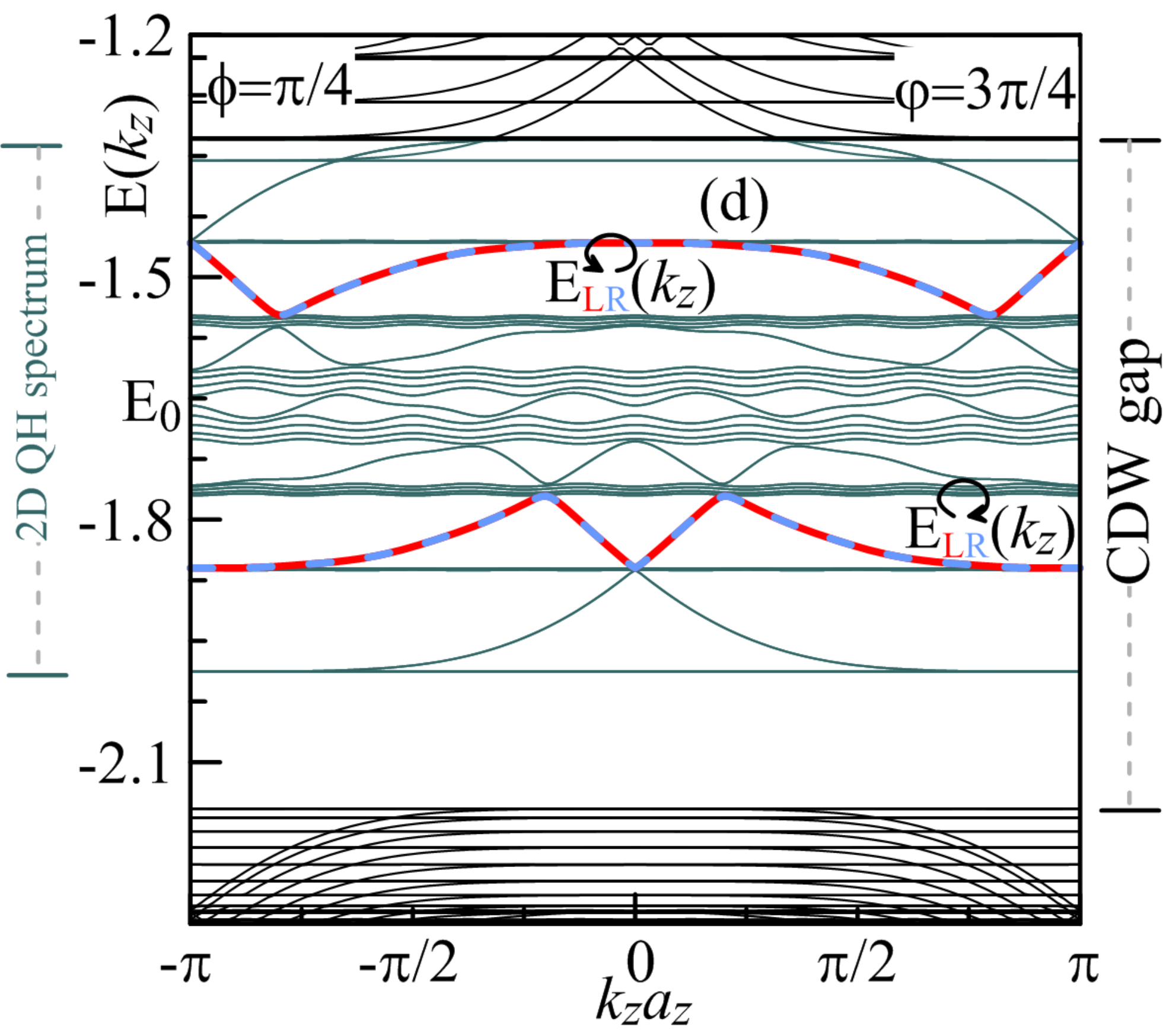}
	\caption{Spectrum of the 3D system shown in Fig.~1(a) of the main text with  periodic boundary conditions in the $z$ direction around the energetically lowest  CDW-induced gap as a function of $k_z$ for $\varphi=3\pi/4$ - same as Fig.~3(a) of the main text but for the filling factor $\nu=2$ with $\phi=\pi/4$. The corresponding doubly degenerate hinge modes $E_{LR}^\sigma(k_z)$ are marked by the red/blue dashed lines. }
	\label{Fig3pi_4_ff2}
\end{figure}

As one can expect the hinge modes with higher QHE filling factors $\nu$ can be supported in our model. Here we show the example for $\nu=2$ for which there are two hinge modes per surface, see Fig.~\ref{Fig3pi_4_ff2}.

\bibliographystyle{unsrt}

\end{document}